\documentclass[amsmath,amssymb,superscriptaddress,nobalancelastpage,aps,longbibliography,twocolumn,showpacs]{revtex4-1}

\usepackage{array,multirow,graphicx}
\usepackage{braket}
\usepackage{epstopdf}
\usepackage{graphicx}
\usepackage{hyperref}
\usepackage{siunitx}
\usepackage{nicefrac}
\usepackage[normalem]{ulem}
\usepackage{varioref}
\usepackage{xcolor}
\usepackage{xfrac}
\usepackage{xr-hyper}

\definecolor{bl}{rgb}{0.0,0.2,0.6}

\hypersetup{colorlinks,linkcolor=blue,urlcolor=blue,citecolor=blue}

\newcommand{\Ca}{Ca$_3$Ru$_2$O$_7$}

\newcommand{\sCa}{Ca$_2$RuO$_4$}

\newcommand{\dxy}{$d_{xy}$}
\newcommand{\dxz}{$d_{xz}$}
\newcommand{\dyz}{$d_{yz}$}

\begin{document}
 \author{M.~Horio}
 \email{mhorio@issp.u-tokyo.ac.jp}
   \affiliation{Physik-Institut, Universit\"{a}t Z\"{u}rich, Winterthurerstrasse 190, CH-8057 Z\"{u}rich, Switzerland}
   
     \author{Q.~Wang}
     \affiliation{Physik-Institut, Universit\"{a}t Z\"{u}rich, Winterthurerstrasse 190, CH-8057 Z\"{u}rich, Switzerland}
     
     \author{V.~Granata}
\affiliation{CNR-SPIN, I-84084 Fisciano, Salerno, Italy}
\affiliation{Dipartimento di Fisica "E.R.~Caianiello", Universit\`{a} di Salerno, I-84084 Fisciano, Salerno, Italy}
     
       \author{K.~P.~Kramer}
     \affiliation{Physik-Institut, Universit\"{a}t Z\"{u}rich, Winterthurerstrasse 190, CH-8057 Z\"{u}rich, Switzerland}
     
 \author{Y.~Sassa}
\affiliation{Department of Physics, Chalmers University of Technology, SE-412 96 G\"{o}teborg, Sweden}
     
       \author{S.~J\"ohr}
     \affiliation{Physik-Institut, Universit\"{a}t Z\"{u}rich, Winterthurerstrasse 190, CH-8057 Z\"{u}rich, Switzerland}
     
       \author{D.~Sutter}
     \affiliation{Physik-Institut, Universit\"{a}t Z\"{u}rich, Winterthurerstrasse 190, CH-8057 Z\"{u}rich, Switzerland}
     
      \author{A.~Bold}
     \affiliation{Physik-Institut, Universit\"{a}t Z\"{u}rich, Winterthurerstrasse 190, CH-8057 Z\"{u}rich, Switzerland}
     
      \author{L.~Das}
     \affiliation{Physik-Institut, Universit\"{a}t Z\"{u}rich, Winterthurerstrasse 190, CH-8057 Z\"{u}rich, Switzerland}
     
     \author{Y.~Xu}
     \affiliation{Physik-Institut, Universit\"{a}t Z\"{u}rich, Winterthurerstrasse 190, CH-8057 Z\"{u}rich, Switzerland}
     
     \author{R.~Frison}
     \affiliation{Center for X-ray Analytics, Swiss Federal Laboratories for Materials Science and Technology (Empa), \"{U}berlandstrasse 129, CH-8600 D\"{u}bendorf, Switzerland}

 \author{R.~Fittipaldi}
\affiliation{CNR-SPIN, I-84084 Fisciano, Salerno, Italy}
\affiliation{Dipartimento di Fisica "E.R.~Caianiello", Universit\`{a} di Salerno, I-84084 Fisciano, Salerno, Italy}

\author{T.~K.~Kim}
\affiliation{Diamond Light Source, Harwell Campus, Didcot, OX11 0DE, United Kingdom}

\author{C.~Cacho}
\affiliation{Diamond Light Source, Harwell Campus, Didcot, OX11 0DE, United Kingdom}

\author{J.~E.~Rault}
\affiliation{Synchrotron SOLEIL, Saint-Aubin-BP 48, F-91192 Gif sur Yvette, France}

\author{P.~Le~F\`{e}vre}
\affiliation{Synchrotron SOLEIL, Saint-Aubin-BP 48, F-91192 Gif sur Yvette, France}

\author{F.~Bertran}
\affiliation{Synchrotron SOLEIL, Saint-Aubin-BP 48, F-91192 Gif sur Yvette, France}

\author{N.~C.~Plumb}
\affiliation{Swiss Light Source, Paul Scherrer Institut, CH-5232 Villigen PSI, Switzerland}
 
\author{M.~Shi}
\affiliation{Swiss Light Source, Paul Scherrer Institut, CH-5232 Villigen PSI, Switzerland}


   \author{A.~Vecchione}
\affiliation{CNR-SPIN, I-84084 Fisciano, Salerno, Italy}
\affiliation{Dipartimento di Fisica "E.R.~Caianiello", Universit\`{a} di Salerno, I-84084 Fisciano, Salerno, Italy}
   
    \author{M.~H.~Fischer}
  \affiliation{Physik-Institut, Universit\"{a}t Z\"{u}rich, Winterthurerstrasse 190, CH-8057 Z\"{u}rich, Switzerland}
   
  \author{J.~Chang}
  \email{johan.chang@physik.uzh.ch}
    \affiliation{Physik-Institut, Universit\"{a}t Z\"{u}rich, Winterthurerstrasse 190, CH-8057 Z\"{u}rich, Switzerland}

    






\title{Electronic reconstruction forming a $C_2$-symmetric Dirac semimetal in Ca$_3$Ru$_2$O$_7$}

\maketitle


  \begin{figure*}[ht!]
 	\begin{center}
 		\includegraphics[width=0.9\textwidth]{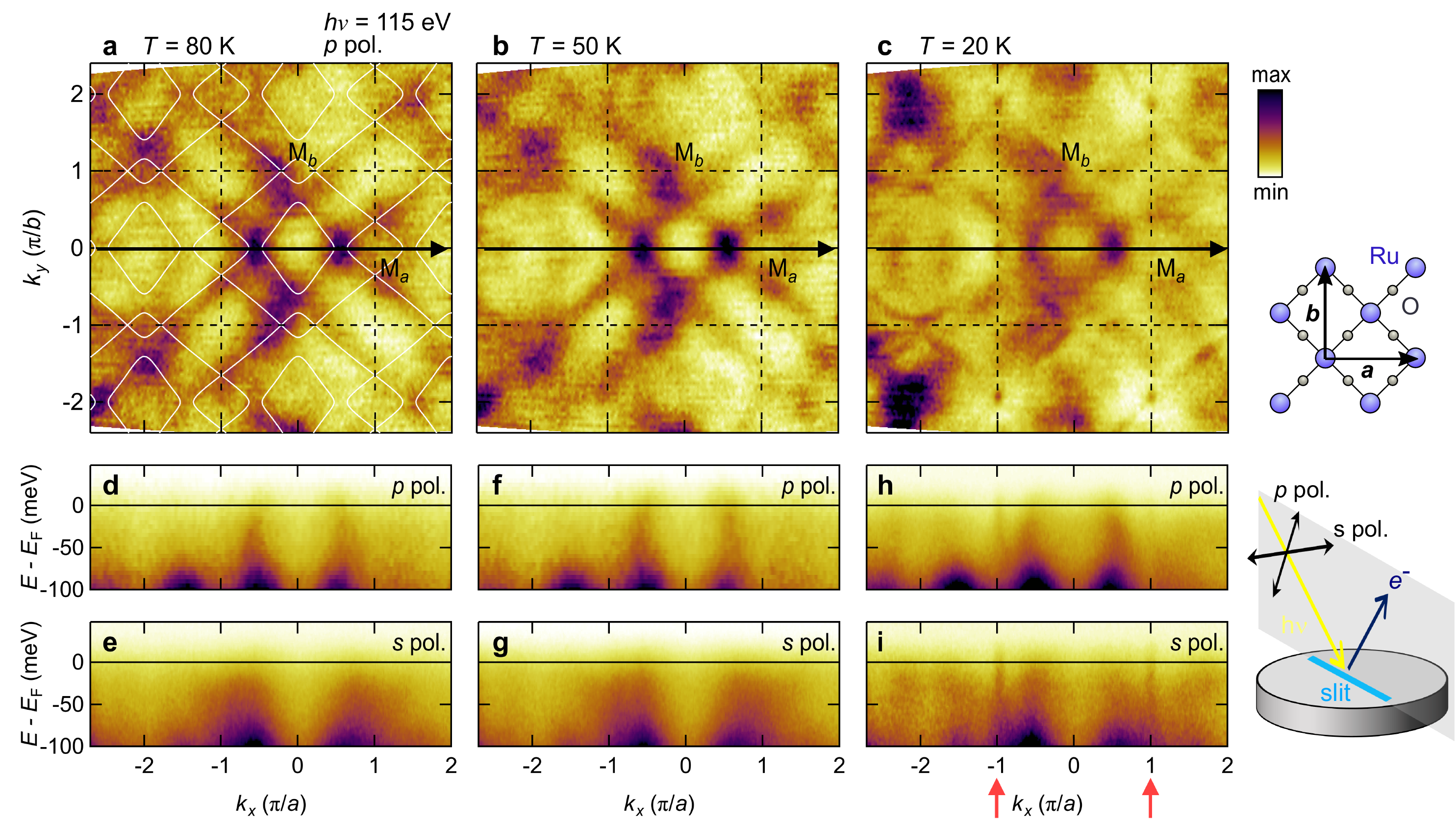}
 	\end{center}
 	\caption{\textbf{Fermi-surface reconstruction in \Ca.} (a)--(c) Fermi surface maps taken at $T=$ 80, 50, and 20~K, respectively. The photoelectron intensities -- displayed using a false color scheme -- are integrated within $E_\mathrm{F} \pm 20$~meV. White lines in (a) represent a  tight-binding model (see Supplementary Note~1 and Supplementary Fig.~1) of the Fermi surface. Vertical and horizontal dashed lines indicate the orthorhombic zone boundaries. The orthorhombic $a$ and $b$ axes are indicated in top right inset. 
 	(d)--(i) Energy distribution maps along the $\Gamma$--M$_a$ direction for incident light polarisations and temperatures as indicated.  
 	Red arrows mark the fast dispersing bands that emerge below $T_\mathrm{s1}=48$~K.  The photoemission mirror plane along with incident light polarisation and the orientation of the electron analyser slit is shown in the bottom inset. With this setting, matrix elements for even-parity orbitals are 
 	expected to be suppressed in the $s$-polarisation channel. 
 	}	
	 	\label{fig:fig1}
 \end{figure*}

\textbf{
Electronic band structures in solids stem from a periodic potential reflecting the structure of either the crystal lattice or an electronic order. 
In the stoichiometric ruthenate Ca$_3$Ru$_2$O$_7$, numerous Fermi surface sensitive probes indicate a low-temperature electronic 
reconstruction. Yet, the causality and the reconstructed band structure remain unsolved.
Here, we show by angle-resolved photoemission spectroscopy,  how in Ca$_3$Ru$_2$O$_7$ a $C_2$-symmetric massive Dirac semimetal is realized through a Brillouin-zone preserving 
electronic reconstruction.
This Dirac semimetal emerges in a two-stage transition upon cooling. The Dirac point and band velocities are consistent with constraints set by quantum oscillation, thermodynamic, and transport experiments, suggesting that 
the complete Fermi surface is resolved. The reconstructed structure -- incompatible with translational-symmetry-breaking density waves -- serves as an important test for band structure calculations of correlated electron systems.}\\[2mm]

\textbf{Introduction:}\\
A Fermi-surface reconstruction refers to the sudden change of the electronic band structure as a function of a tuning parameter. As most electronic properties are governed by electrons in the vicinity of the
Fermi level, a change of the Fermi-surface topology can have profound ramifications. Those reconstructions that are not obviously linked to a symmetry change of the crystal structure are of particular interest.
Common triggers of Fermi-surface reconstructions are translational-symmetry breaking spin- or charge-density waves.
Typically, this reduction of symmetry and the resulting folding of the Brillouin zone lead large Fermi-surface contours to split into smaller pockets. Once reconstructed, 
it is, however, often difficult to identify the Fermi surface structure. The high-temperature superconductor YBa$_2$Cu$_3$O$_{7-x}$ is a good example of this. In the underdoped regime, charge-density-wave order~\cite{WuNature2011,GhiringhelliScience2012,ChangNatPhys2012} clearly reconstructs the Fermi surface. Although quantum oscillations~\cite{Doiron-LeyraudNature2007,SebastianAR2015} and transport~\cite{LeBoeufNature2007,ChangPRL2010} experiments have revealed the existence of an electron pocket, the reconstructed Fermi-surface topology is, despite strong effort, still not clarified~\cite{HossainNatPhys2008}. Another prominent example is the orthorhombic bilayer ruthenate \Ca. An initial A-type antiferromagnetic (AFM) order setting in at $T_\mathrm{N} = 56$~K~\cite{YoshidaPRB2005} has no
significant impact on the 
transport properties~\cite{YoshidaPRB2004}. 
A spin reorientation from the orthorhombic $a$- 
to the $b$-axis 
direction~\cite{BaoPRL2008,BohnenbuckPRB2008} occurs at $T_\mathrm{s1}=48$~K.
While this lattice-space-group preserving reorientation~\cite{YoshidaPRB2005} 
naively might appear to be of minor consequence, the transition at $T_{s1}$ = 48 K marks a dramatic electronic transformation.

Across this transformation, the Seebeck coefficient undergoes a sharp sign change taking large negative values below the transition temperature~\cite{IwataJMMM2007,XingPRB2018}. $T_\mathrm{s1}$ is also the onset of
in-plane anisotropic transport properties~\cite{XingPRB2018}.
Although transport and thermodynamic experiments 
provide information 
about Fermi surface area and electronic masses~\cite{CaOPRB2003,KikugawaJPSJ2010}, the complete reconstructed Fermi surface 
has so far remained 
undetermined. Previously reported angle-resolved photoemission spectroscopy (ARPES) data has revealed the existence of a boomerang-shaped Fermi surface~\cite{BaumbergerPRL2006}. Obviously, a single pocket is insufficient to explain 
the observed ambipolar electronic properties~\cite{XingPRB2018}. Although a density-wave state has never been identified~\cite{BohnenbuckPRB2008}, a common assumption is that a translational-symmetry-breaking order reconstructs the Fermi surface into multiple sheets. However, as long as the complete Fermi surface and its orbital composition remain unidentified, so do 
the reconstructing mechanism.

Here, we provide by direct ARPES experiments the complete Fermi surface structure of \Ca\ across the 
electronic reconstruction. Above the reconstructing temperature, the low-energy electronic structure resembles that of a strongly correlated metal consistent with its orthorhombic crystal structure. The reconstructed Fermi surface by contrast consists of a small electron pocket formed by massive Dirac fermions along the short-axis orthorhombic zone boundary and a boomerang-like hole pocket in vicinity to the long-axis zone boundary. 
As the
orthorhombic order parameter $|a-b|/(a+b)$ 
remains essentially unchanged across the reconstruction~\cite{YoshidaPRB2005}, the sudden emergence of such a dramatic Fermi surface anisotropy is unexpected.
We furthermore demonstrate that the Fermi surface transformation appears in two steps. The
anisotropic zone-boundary Fermi surfaces first appear below $T_\mathrm{s1}=48$~K and eventually the Dirac fermions settle into the low-temperature structure below $T_\mathrm{s2}=30$~K. Throughout this temperature evolution, no signature of Brillouin-zone folding is identified, excluding 
translational-symmetry breaking
density-wave/orbital orders as the origin of the phase transition at $T_\mathrm{s1}=48$~K. We argue that the reconstructed Fermi surface should be understood from the \dxz\ and \dyz\ orbitals whereas the \dxy\ sector is not crossing the Fermi level in the reconstructed phase. The revelation of the complete Fermi surface reconstruction provides an ideal test-bed for ab-initio band structure calculations beyond density-functional-theory concepts.   \\[2mm]

 \begin{figure*}[ht!]
 	\begin{center}
 		\includegraphics[width=0.9\textwidth]{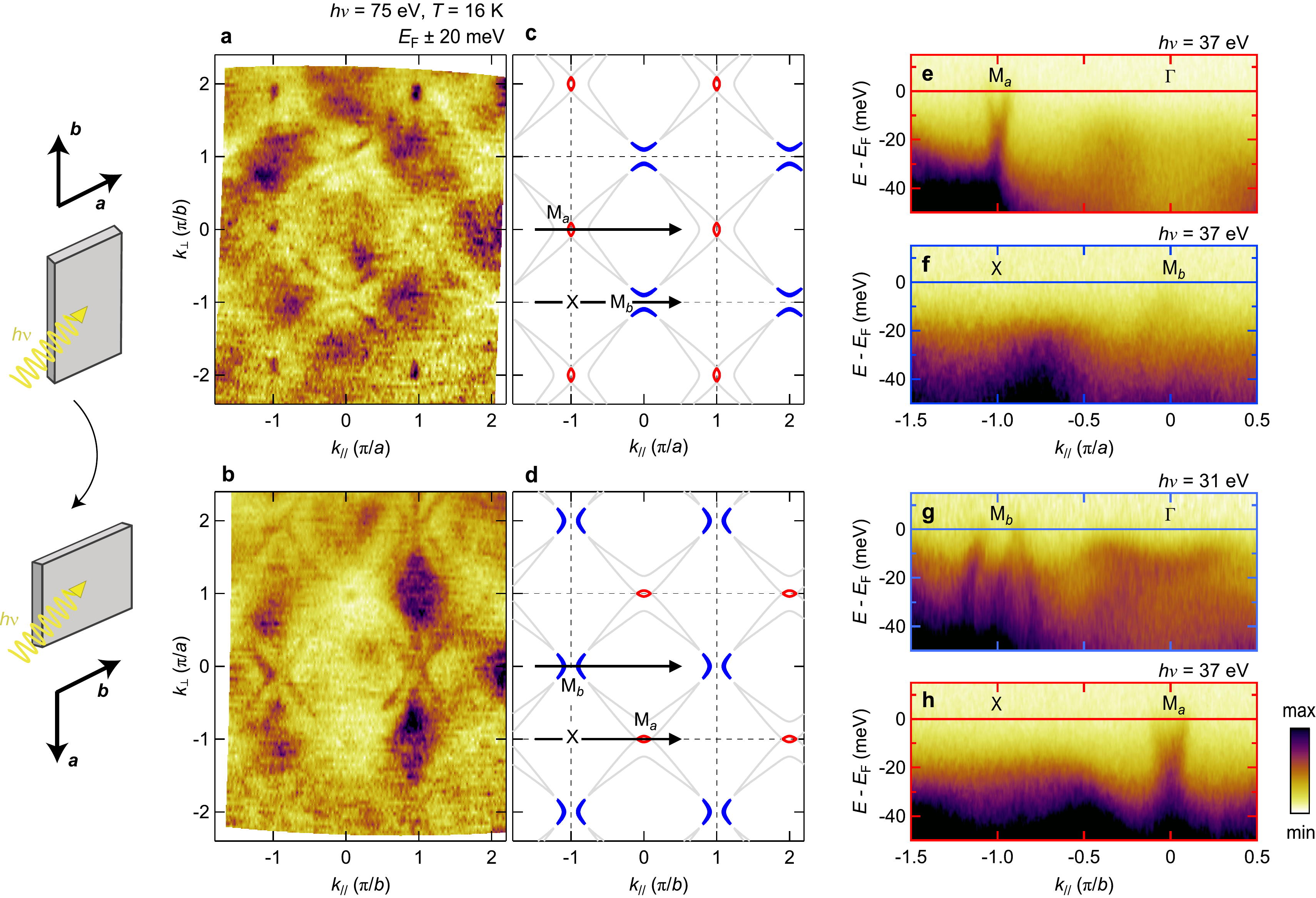}
 	\end{center}
 	\caption{\textbf{Breaking of four-fold rotational symmetry.} (a),(b) Fermi surface maps recorded at $T = 16$~K ($h\nu = 75$~eV) for two sample azimuthal angles
 	 that are 90$^\circ$ apart as indicated in the schematics. The azimuthal angle rotation was operated \textit{in-situ} and hence the sample surface is identical. $k_{\parallel}$ ($k_{\perp}$) represents momentum parallel (perpendicular) to the electron-analyser slit. The spectral intensity was integrated within $E_\mathrm{F} \pm 20$~meV, washing-out spectral gaps within the integration window. (c),(d) Fermi surfaces from the tight-binding model~(see Supplementary note~1). The sheets not observed in the experiment are indicated in gray. (e)--(h) Energy distribution maps along 
 	M$_a$--$\Gamma$, X--M$_b$, M$_b$--$\Gamma$, and X--M$_a$
 	as shown in panels (c) and (d). The energy distribution maps were recorded at $h\nu = 37$~eV except for (g) where $h\nu = 31$~eV incident light was selected to optimise the photoelectron matrix element. Systematic photon-energy dependence between 30 and 40~eV is presented in Supplementary Figs.~2 and 3.  The band structure along M$_a$--$\Gamma$ (X--M$_a$) and M$_b$--$\Gamma$ (X--M$_b$) is inequivalent.} 
	 	\label{fig:fig2}
 \end{figure*}

\textbf{Results:}\\
\textbf{High-temperature state:} The Fermi surface and low-energy electronic structure of the \Ca\ normal state -- above the N\'eel temperature $T_\mathrm{N}=56$~K -- are presented in Figs.~\ref{fig:fig1}(a), (d), and (e). The orthorhombic zone boundary is indicated by black dashed lines in Fig.~\ref{fig:fig1}(a). All quasiparticle dispersions are broad irrespective of whether linear $p$- or $s$-polarised light is used. 
Part of the Fermi surface consists of 
straight sectors running diagonally through the orthorhombic Brillouin zone. 
This quasi-one-dimensional structure remains essentially unchanged across the N\'eel transition at $T_\mathrm{N}=56$~K [see Figs.~\ref{fig:fig1}(b), (f), and (g)]. Furthermore, the orthorhombic zone boundary points $\mathrm{M}_a=(\pm\pi/a, 0)$ and $\mathrm{M}_b=(0, \pm\pi/b)$ are virtually indistinguishable (See also Supplementary Fig.~1).\\[1mm]

\begin{figure*}[ht!]
 	\begin{center}
 		\includegraphics[width=1\textwidth]{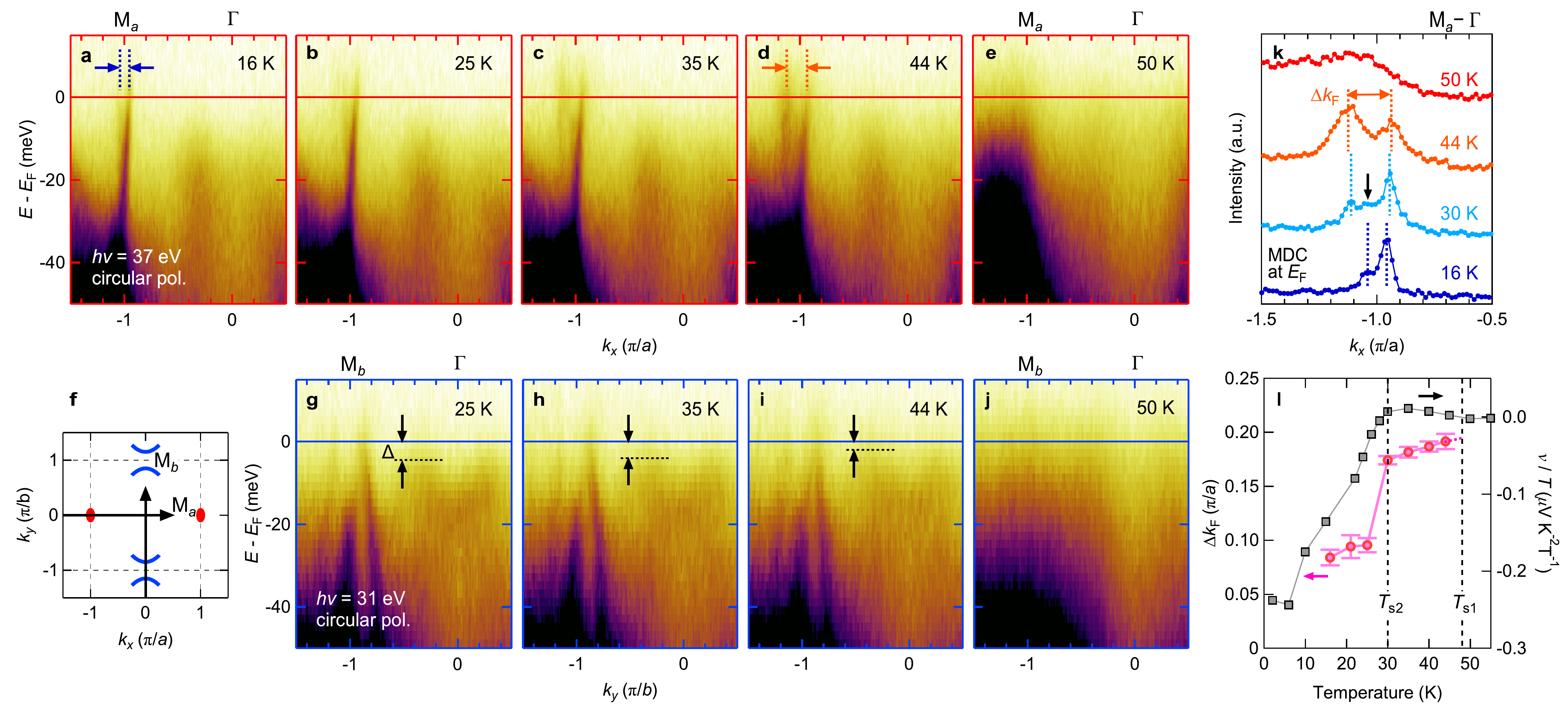}
 	\end{center}
 	\caption{\textbf{Two-stage temperature evolution of the band structure.} (a)--(j) Energy distribution maps 
 	along M$_a$--$\Gamma$ (top panels) and M$_b$--$\Gamma$ (bottom panels)
 	for temperatures as indicated.
 	The gap $\Delta$ of the electron pocket around $\Gamma$ is indicated in (g)--(i). For detailed analysis, see Supplementary Fig.~4.
 	(f) Schematics of the Fermi surface and the high-symmetry cuts used in (a)--(e) and (g)--(j). (k) MDCs at $E_\mathrm{F}$ (integrated within $\pm 3$~meV) 
 	for temperatures as indicated. Vertical dashed lines define MDC peak maxima. A clear difference in the peak separation, $\Delta k_\mathrm{F}$, is found across 30~K. At 30~K, a three-peak structure is found and indicated by the black arrow and the two dashed lines. 
 	(l) $\Delta k_\mathrm{F}$ plotted as a function of temperature. The error bars represent $3\sigma$ of the fitting with $\sigma$ being the standard deviation. For comparison, the Nernst coefficient $\nu$ \cite{XingPRB2018} is plotted as $\nu/T$ versus $T$. Both experiments suggest an electronic transformation across $T=30$~K.}
	 	\label{fig:fig3}
 \end{figure*}

\textbf{Fermi surface anisotropy:} Across the structural transition at $T_\mathrm{s1} = 48$~K, however, the electronic structure undergoes a dramatic reconstruction. This is evidenced by the emergence of a fast dispersing band and a tiny Fermi surface around $\mathrm{M}_a$
-- see Figs.~\ref{fig:fig1}(c), (h), and (i). Remarkably, this small Fermi surface sheet is 
absent at $\mathrm{M}_b$.
Instead, as previously reported~\cite{BaumbergerPRL2006}, boomerang-like Fermi surface sheets are found around the $\mathrm{M}_b$ point. 
Therefore, in contrast to $T>T_\mathrm{s1}$,
the low-temperature structure 
appears highly anisotropic featuring 
different Fermi surface topology
around M$_a$ and M$_b$. This Fermi surface reconstruction appears without change of the crystal lattice space group and with minute ($\sim 1$\%) reduction 
of the orthorhombic order parameter~\cite{YoshidaPRB2005}. 


To exclude the possibility that this $C_2$ symmetry is an artifact of photoionization-matrix-element effects, we follow a standard measurement protocol~\cite{YiPNAS2011,WatsonNJP2017}. That is to carry out Fermi surface mappings with azimuthal angles differing from each other by 90$^\circ$ [see Figs.~\ref{fig:fig2}(a) and (b)]. 
\textit{In-situ} azimuthal rotation implies that the Fermi surface maps in Figs.~\ref{fig:fig2}(a) and (b) are from the same surface.
Here, $k_{\parallel}$ ($k_\perp$) on the horizontal (vertical) axis represents the momentum parallel (perpendicular) to the electron-analyser slit. 
The electronic structure with a tiny Fermi pocket around the $\mathrm{M}_a$ point and boomerang-like features near $\mathrm{M}_b$ tracks the azimuthal rotation -- see Figs.~\ref{fig:fig2}(a)--(d).
The $C_2$ symmetric electronic structure is also revealed by the band dispersions. Along the M$_a$--$\Gamma$ and M$_b$--$\Gamma$ 
directions, the band curvature around M$_a$ and M$_b$ are clearly different [Figs.~\ref{fig:fig2}(e) and (g)]. An electron pocket is formed around M$_a$ whereas two hole-like pockets 
are found on each side of M$_b$. In a similar fashion, dispersions along the M$_a$--X and M$_b$--X directions are inequivalent [Figs.~\ref{fig:fig2}(f) and (h)]. Electron-like band curvature is found around $\mathrm{M}_a$ whereas no Fermi crossing is observed along M$_b$--X. 
These results 
exclude matrix-element effects as the source of the observed 
anisotropy.\\[1mm] 

\textbf{Two-stage Fermi surface reconstruction:}
Tracking the temperature dependence of the band structure reveals two electronic temperature scales. 
The electronic band structure along M$_a$--$\Gamma$ and M$_b$--$\Gamma$ is shown for temperatures going from 16~K to 50~K.
 Above $T_\mathrm{s1}=48$~K [Figs.~\ref{fig:fig3}(e), (j) and Fig.~\ref{fig:fig1}], all bands appear with broad line-shapes. Once cooled below $T_\mathrm{s1}$, well-defined bands around the $\mathrm{M}_a$ and $\mathrm{M}_b$ points emerge [see Figs.~\ref{fig:fig3}(d), (i), and (k)]. 
The appearance of another electron-like band around the $\Gamma$ point is accompanied by a gap $\Delta$ opening below $T_{s1}$ [see Figs.~\ref{fig:fig3}(g)-(i) and  Supplementary Fig.~4].
 The band structures around $\mathrm{M}_a$ and $\mathrm{M}_b$ are inequivalent not only in terms of curvature but also in terms of temperature dependence. The $\mathrm{M}_a$--$\Gamma$ band dispersion is temperature dependent whereas the corresponding structure around $\mathrm{M}_b$ is virtually insensitive to temperature.
 Examining the $\mathrm{M}_a$--$\Gamma$ direction, two inequivalent bands with different Fermi momenta are observed for $30<T<48$~K
 whereas only a single set of bands is resolved for $T<30$~K -- see Fig.~\ref{fig:fig3}. The two bands around $M_a$ display asymmetric matrix elements. Momentum distribution curves (MDCs) at  $E_\mathrm{F}$ and $30<T<48$~K are therefore not 
 symmetric around M$_a$. Upon cooling below 30~K, the single electron pocket around $M_a$ display 
 symmetric Fermi momenta $k_\mathrm{F}$ despite the asymmetric matrix elements.  To illustrate the transition between the two- and single-band situation, we define $\Delta k_\mathrm{F}$ as the reciprocal-space distance between the two MDC-intensity maxima. Across $T_\mathrm{s2}=30$~K, $\Delta k_\mathrm{F}$ drops by a factor of two. This observation is independent of incident-light polarisation (see Supplementary Figs.~5 and 6). On the other hand, the gap $\Delta$ evolves smoothly across $T_\mathrm{s2}$ [see Figs.~\ref{fig:fig3}(g)--(i) and Supplementary Fig.~4], suggesting that the states around $\Gamma$ are not involved in this transition.
 Although $T_\mathrm{s2}=30$~K remains to be identified as a thermodynamic temperature scale, it does coincide with the onset of a strong negative Nernst effect response ~\cite{XingPRB2018} [see Fig.~\ref{fig:fig3}(l)]. We also notice that the reduced Fermi momenta $\Delta k_\mathrm{F}$ below $T_\mathrm{s2}$ is consistent with an increasing Nernst effect $\nu/T \propto \mu / E_\mathrm{F}$ where $\mu$ is the electron mobility and $E_\mathrm{F}$ the Fermi energy~\cite{Behnia2009}, as lower $\Delta k_\mathrm{F}$ implies a smaller Fermi energy. 
 The low-temperature Fermi surface thus emerges as a result of two reconstructions. First below
 $T_\mathrm{s1}=48$~K, a fast dispersing band appears around $\mathrm{M}_a$ and $\mathrm{M}_b$ with a gap opening for other bands. 
 Next, the band dispersion along the M$_a$--$\Gamma$ direction undergoes a second transformation across $T_\mathrm{s2}=30$~K.\\[1mm]

\textbf{Low-temperature electronic structure:} 
With the exception of 
the 
features around the M$_a$ and M$_b$ points, all other bands are not crossing the Fermi level for $T < T_\mathrm{s2}$ --
see Figs.~\ref{fig:fig2}(e)--(h). 
Around the $\mathrm{M}_b$ point, two hole-like bands -- forming an M-shaped structure -- are found [Fig.~\ref{fig:fig2}(g)]. While the hole-like band touches $E_\mathrm{F}$ along the M$_b$--$\Gamma$ direction [Fig.~\ref{fig:fig2}(g)], the band top sinks below $E_\mathrm{F}$ upon moving away from it by $\sim 0.1 \pi/b$. (See Supplementary Fig.~7) consistent with a previous report~\cite{BaumbergerPRL2006}. The boomerang-like feature thus forms a closed hole-like Fermi surface. 

\begin{figure*}[ht!]
 	\begin{center}
 		\includegraphics[width=0.85\textwidth]{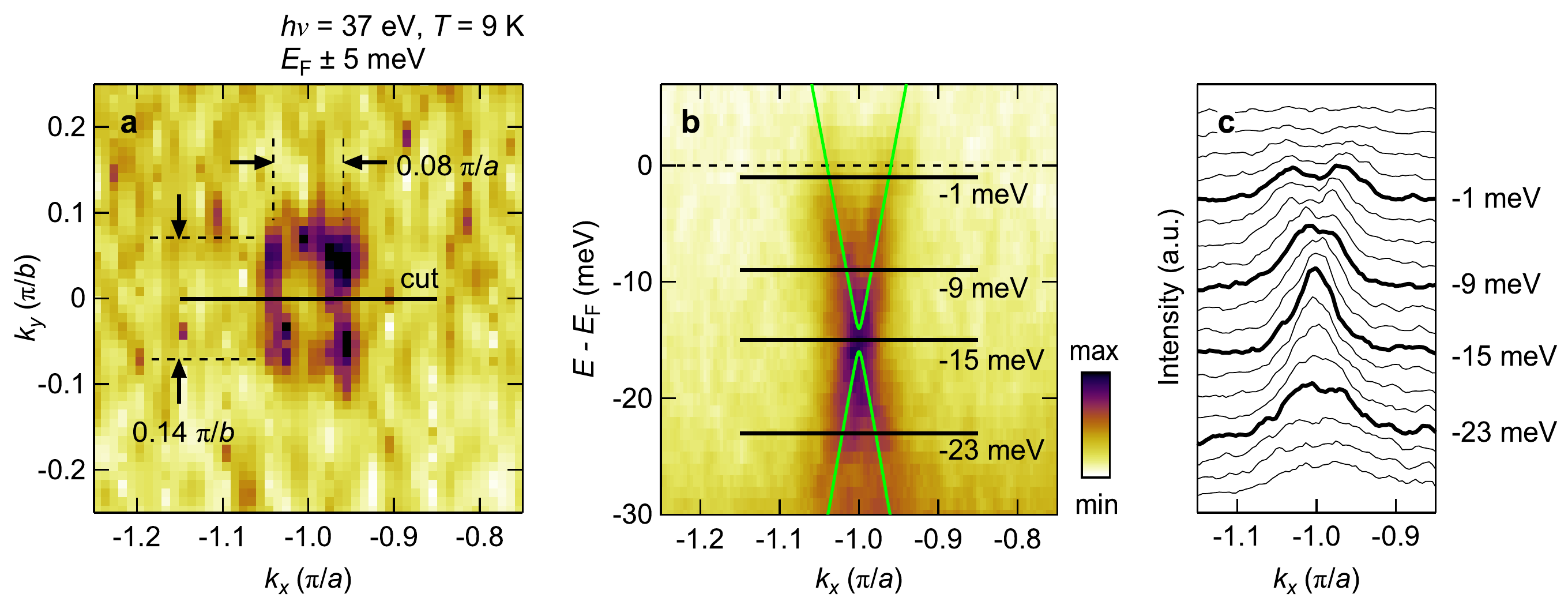}
 	\end{center}
 	\caption{\textbf{Dirac cone structure in Ca$_3$Ru$_2$O$_7$} (a) High-resolution Fermi surface map around the M$_a$ point. (b) Energy distribution map along the M$_a$--$\Gamma$ direction evidencing a Dirac cone structure. Crystal/magnetic symmetry imposes a finite mass as illustrated by green dispersions reproduced by the tight-binding model (see Supplementary Note~1). (c) MDCs, extracted from (b), at binding energies as indicated.
 	}	
	 	\label{fig:fig4}
 \end{figure*}

Around the $\mathrm{M}_a$ point, the electron-like Fermi surface pocket is revealed by a high-resolution map in Fig.~\ref{fig:fig4}(a). The 
electron pocket is elliptical with $k_\mathrm{F}^{a}= 0.04 \pi/a$ and $k_\mathrm{F}^{b}= 0.07 \pi/b$ along the 
M$_a$--$\Gamma$ and M$_a$--X directions, respectively.
The Fermi surface area $A_\mathrm{FS}=\pi k_\mathrm{F}^a k_\mathrm{F}^b$ corresponds to 
0.23\% of the orthorhombic Brillouin zone.
Inspecting the band dispersion along the M$_a$--$\Gamma$ direction reveals a Dirac-cone structure with the Dirac point placed about $E_\mathrm{D}=15$~meV below $E_\mathrm{F}$ [Fig.~\ref{fig:fig4}(b)]. 
The two-peak MDC profile found at $E_\mathrm{F}$ merges into a single peak at $E_\mathrm{D}$ and then splits again below $E_\mathrm{D}$ [see Fig.~\ref{fig:fig4}(c)]. This MDC analysis estimates
a linear Fermi velocity of $v_\mathrm{F}^a=0.62$~eV\AA\ (95~km/s) and $v_\mathrm{F}^b=0.37$~eV\AA\ (57~km/s) 
(see Supplementary Fig.~8). Our results thus suggest that Ca$_3$Ru$_2$O$_7$ at low temperatures is a highly anisotropic Dirac semimetal.\\[2mm]

\textbf{Discussion:}\\
The low-temperature resistivity anisotropy in \Ca, is as large as $\rho_c/\rho_{ab}\sim 1000$~\cite{YoshidaPRB2004}. It is, therefore, reasonable to consider the  Fermi surface to be two-dimensional. This implies that quantum oscillation and ARPES experiments are directly comparable.
Onsager's relation~\cite{SebastianAR2015} links directly Fermi surface areas to quantum oscillation frequencies $F=\Phi_0 A_\mathrm{FS} / (2\pi^2)$ where $\Phi_0$ is the flux quantum.
The low-temperature electron pocket around $\mathrm{M}_a$
corresponds to $F=34$~T in 
agreement with 
observed quantum oscillation frequencies 28--43~T~\cite{CaOPRB2003,KikugawaJPSJ2010,XingPRB2018}.
The hole-like boomerang structure comprises a Fermi surface area that is too small to be quantified accurately by our ARPES experiments. However, it should produce a low-frequency
quantum oscillation. Indeed, a frequency corresponding to 0.07\% of the Brillouin zone or about $1/3$ of the electron pocket has been reported ~\cite{CaOPRB2003}.
It is therefore conceivable that the electron and hole Fermi pockets reported here are those responsible for the quantum oscillations. Our ARPES work unveils the band curvature and position of these pockets within the Brillouin zone.

Combining ARPES and quantum oscillation experiments allow direct comparison of the effective electronic mass $m^* = \frac{\hbar^2}{2\pi} \frac{\partial A_\mathrm{FS}}{\partial \varepsilon}$~\cite{RourkeNJP2010,HorioPRL2018} on the electron pocket.
Lifshitz-Kosevich analysis of the $\sim35$~T quantum-oscillation frequency yield $m_e^*=0.6 m_e$~\cite{KikugawaJPSJ2010}, where $m_e$ is the free electron mass. Assuming a parabolic band dispersion 
$m_e^*=\hbar^2 k_\mathrm{F}^a k_\mathrm{F}^b /2\varepsilon_\mathrm{F}$ where 
 $\varepsilon_\mathrm{F}=15$~meV is the Fermi energy, the effective mass 
 $m_e^*= 0.25 m_e$ is significantly lower than
that inferred from quantum-oscillation experiments. 
A linear band dispersion $E_k=v_\mathrm{F} k$ provides a much better agreement 
$m_e^*=\varepsilon_\mathrm{F}/v_\mathrm{F}^a v_\mathrm{F}^b = \hbar^2 k_\mathrm{F}^a k_\mathrm{F}^b / \varepsilon_\mathrm{F} =0.49m_e$.
This fact reinforces the interpretation of Dirac fermions around $\mathrm{M}_a$.
The boomerang band along $\Gamma$--$\mathrm{M}_b$ has comparable Fermi velocity to that of the electron pocket along $\Gamma$--$\mathrm{M}_a$. 
Estimation of the Fermi energy from linear extrapolation of the M-shaped band dispersion yields $\varepsilon_\mathrm{F}\sim5$~meV, which is about three times smaller than that obtained for the electron pocket.
As the hole pocket area -- according to quantum-oscillation measurements~\cite{CaOPRB2003} -- is also about 3 times smaller than the electron sheet, 
we estimate the hole-like carriers to have a comparable  effective mass of $m_h^*=0.49m_e$. With two hole and one electron pocket per Brillouin zone, a Sommerfeld constant of 
$\gamma\approx2.1$~mJ/(mol K$^2$) is found, with mol refering to one formula unit. Here, we assumed two-dimensional band dispersions without bi-layer splitting and used the two-dimensional expression $\gamma = \pi N_\mathrm{A} k_\mathrm{B}^2 ab/3\hbar^2 \sum_i m_i$~\cite{KikugawaJPSJ2010}, where $N_\mathrm{A}$ is the Avogadro constant, $k_\mathrm{B}$ is the Boltzmann constant, $\hbar$ is the reduced Planck constant, and $m_i$ is the effective mass. In addition, spin polarisation within the RuO$_2$ plane was employed to treat the A-type AFM order. As our estimate is in reasonable agreement with the value $\gamma\approx2.8$--3.4 mJ mol$^{-1}$ K$^{-2}$) obtained by specific heat experiments~\cite{YoshidaPRB2004,KikugawaJPSJ2010}, 
we conclude that our experiments reveal the entire 
bulk Fermi surface.

The two-stage transformation of the electronic structure has a clear impact on all transport coefficients. A remarkable increase of in- and out-of-plane resistivity appears across $T_\mathrm{s1}=48$~K~\cite{YoshidaPRB2004,KikugawaJPSJ2010}. Simultaneously, the Seebeck coefficient changes sign going from weak positive to large negative values across $T_\mathrm{s1}$~\cite{XingPRB2018}. Although less sharp, the Hall coefficient also changes sign (from positive to negative) across $T_\mathrm{s1}$. While the Hall coefficient takes increasingly large negative values~\cite{YoshidaJPSJ2007,XingPRB2018}, the Seebeck coefficient displays a complicated temperature dependence that in addition is different along the $a$ and $b$ directions~\cite{XingPRB2018}. 
This temperature dependence is a typical signature of ambipolar transport behaviour, where both electron- and hole-like carriers are contributing~\cite{BelPRL2003}. Furthermore, the low-temperature Hall coefficient 
$R_\mathrm{H}$ that (in different studies) ranges from $-0.5 \times 10^{-7}$~m$^3$ C$^{-1}$ to $-1.4 \times 10^{-7}$~m$^3$/C~\cite{YoshidaJPSJ2007,KikugawaJPSJ2010,XingPRB2018} cannot be explained by the electron pocket that alone should generate $R_\mathrm{H}=-1/(n_e e)=-8.0 \times 10^{-7}$~m$^3$/C. 
Using the combined ARPES and quantum-oscillation knowledge that $n_e=7.8 \times 10^{18}$~cm$^{-3}$ and 
$n_h\approx$ $2 \times n_e/3$, a two band model~\cite{RourkePRB2010} yields 
$R_\mathrm{H}=(2\alpha^2/3-1)/(n_e e)(2\alpha/3+1)^2$ where $\alpha=\mu_h/\mu_e$ is the mobility ratio between electrons and holes. The exact experimental values of $R_\mathrm{H}(T=0)$ imply that 
$\mu_h \approx$ 0.9--1.1 $\times \mu_e$ and 
$\mu_e\approx |R_\mathrm{H}|/\rho_{xx}=0.1$~T$^{-1}$. We thus infer that in the $T \to 0$ limit electron- and hole-like carriers have comparable mobility that in turn generate 
the ambipolar transport properties.

Having established the existence of small electron pockets with linear dispersion around the M$_a$ point, a question arises whether these excitations are massless Dirac fermions or whether they possess a finite mass at M$_a$. While the question cannot be definitively answered from the experimental data due to the finite energy resolution, we discuss here implications from the crystal symmetry. \Ca\ has the space group $Bb2_{1}m$ (No.~36)~\cite{YoshidaPRB2005}. For our purpose, it is sufficient to focus on a single bi-layer. The point group of such a bi-layer is $C_{2v}$ with a mirror plane between the two layers, as well as a glide plane perpendicular to the mirror and a two-fold screw axis along the crystalline $b$ axis (the longer in-plane axis). Together with time-reversal symmetry (TRS) in the paramagnetic state, this imposes a Kramer's degeneracy along M$_b$--X in the Brillouin zone. Futhermore, TRS imposes Kramer's pairs at the M$_a$ and $\Gamma$ point.

When TRS is broken in the A-type AFM phase~\cite{BaoPRL2008}, the generating point group of the bi-layer is reduced to $C_{2v}$ ($C_s$) for the AFM-$a$ phase and $C_{2v}$ ($C_2$) for the AFM-$b$ phase. Here, the notation $\mathcal{G}$ ($\mathcal{G}'$) denotes the generating point group $\mathcal{G}$ with $\mathcal{G}'$ the subgroup of elements that do not have to be combined with TRS. 
While the Kramer's degeneracy is preserved along M$_b$--X, the one at the M$_a$ and $\Gamma$ point is lifted. The Dirac fermions at M$_a$ thus possess a finite mass, in other words the bands hybridize as schematically illustrated in Fig.~\ref{fig:fig4}(b).

Finally, we can reproduce key features of the low-temperature semimetallic band structure employing a tight-binding model of the Ru $t_{2g}$ orbitals (see Supplementary Note~1, Supplementary Figs.~1 and 9).  We restrict our model to the Ru \dxz\ and \dyz\ orbitals in an effective single-layer model, for two reasons. Firstly, the ``one-dimensional'' nature of the high-temperature ($T>48$~K) Fermi surface resembles 
the \dxz, \dyz\ dominated $\alpha$ and $\beta$ bands of other ruthenates
~\cite{DamascelliPRL00_full,TamaiPRL2008}. Secondly, the matrix-element effect of the electron pocket around M$_a$ is  incompatible with the expectation of selection rules for the \dxy\ orbital character [see Figs.~1(h) and (i)]. Our simple
 model faithfully reproduces the Fermi surface in the normal state 
[see Fig.~\ref{fig:fig1}(a) and Supplementary Fig.~1].
Importantly, a rigid band shift in the \dxz/\dyz\ sector, as expected due to the $c$-axis compression at $T_\mathrm{s1}$~\cite{YoshidaPRB2005}, yields elliptical electron pockets with linear dispersion around the M$_a$ point and a hole-like boomerang structure around the M$_b$ point [Figs.~\ref{fig:fig2}(c) and (d)]. This \dxz/\dyz\ band shift implies a change of orbital polarisation to respect the global charge balance. Finally, the Brillouin-zone folding due to the screw-axis opens a gap around the M$_a$ Dirac point [see green lines Fig.~\ref{fig:fig4}(b)]. Due to the small gap size and Dirac point distance from the Fermi level, this gap is irrelevant for transport and thermodynamic measurements.
While our tight-binding model based on the Ru $t_{2g}$ orbitals is too simplistic to capture all the features and does not include the actual electronic instability, it reproduces the most salient features of both the high- and low-temperature dispersions. We thus conclude that the
low-temperature low-energy band structure stems primarily from
the \dxz\ and \dyz\ Ru orbitals.

A fundamental remaining question links to the triggering mechanism  
that induces the Dirac semimetal.
Specific heat suggests that the phase transition at $T_\mathrm{s1}$ involves a large entropy change~\cite{YoshidaPRB2004} and unlike other layered ruthenates, the ground state is a low density-of-state semimetal.
It has been argued that the reorientation of the magnetic moments alone can not account for this large entropy change.
Upon cooling, an energy gain of the system is manifested 
by an electronic reconstruction that opens a gap leaving only small Fermi surface pockets around the zone boundaries. 
Most likely, this Fermi surface reconstruction is triggered by an electronic mechanism. Density-wave orders breaking translation symmetry are, however, excluded since the reconstruction preserves the original Brillouin-zone boundaries. This leads us to speculate alternative scenarios, with electron correlations likely involved in some way.
If so, the situation resembles that of the single-layer counterpart \sCa\ where the instability toward a Mott-insulating state triggers a large $c$-axis lattice contraction~\cite{NakatsujiPRL00,FriedtPRB2001,SutterNatCom2017}.
Indeed, a $c$-axis lattice contraction is found across the first (48~K) transition though this effect is much less pronounced in \Ca~\cite{YoshidaPRB2005}.
Alternatively, it has been proposed that \Ca\ hosts magnetic anapole order~\cite{Thole2018,LoveseyPRB2019}. This would connect \Ca\ with  hidden order problems in the sense that it is very difficult to demonstrate experimentally. 

Note added after completion of this work: A recent complementary ARPES study~\cite{Markovic2020} conducted at $T \geqq 30$~K suggested using DFT calculations including Rashba-type spin-orbit coupling that the electronic reconstruction across $T=48$~K can be understood from the magnetic-moment reorientation alone without the need for additional hidden order.\\

\textbf{Methods:}\\
\textbf{Sample characterisation:}\\
High quality single crystals of \Ca\ were grown by floating zone technique~\cite{YoshidaPRB2004}. The electronic transition at $T_\mathrm{s1}=48$~K was checked by thermopower measurements (see Supplementary Fig.~10) and found in agreement with existing literatures~\cite{IwataJMMM2007,XingPRB2018}. 
Detwinning of orthorhombic domains was achieved with a thermo-mechanical device~\cite{Burkhardt1995} and monitored by polarised light microscopy. The resulting mono domain constitutes 99\% (or more) of the sample volume according to x-ray diffraction measurements (see Supplementary Fig.~10).

\textbf{ARPES experiments:}\\
ARPES experiments were carried out at the SIS~\cite{SIS}, CASSIOPEE, and I05~\cite{HoeschRevSciInst2017_full} beamlines of the Swiss Light Source (SLS), SOLEIL synchrotron, and Diamond Light Source, respectively. Pristine surfaces were obtained by top-post cleaving at $T > T_\mathrm{s1}$ (80~K). 
Incident photons $h\nu = 31-$115~eV, providing high in-plane and modest out-of-plane~\cite{HorioPRL2018} momentum resolution, were used for this study.
Consistent results were obtained on different crystals and upon cooling and heating through the critical temperature $T_\mathrm{s1}=48$~K below which the electronic structure is reconstructed. ARPES data are presented using 
orthorhombic notation with lattice parameters $a=5.37$~\AA\ and $b=5.54$~\AA.\\

\textbf{Data availability} \\
The data that support the findings of this study are available from the corresponding author upon reasonable request. \\

\textbf{Acknowledgements:}\\
We thank M.~Hoesch for fruitful discussions. M.H., Q.W., K.P.K., D.S., Y.X., and J.C. acknowledge support by the Swiss National Science Foundation. Y.S. is funded by the Swedish Research Council (VR) with a Starting Grant (Dnr. 2017-05078). ARPES measurements were carried out at the SIS, CASSIOPEE, and I05 beamlines of the Swiss Light Source, SOLEIL synchrotron, and Diamond Light Source, respectively. We acknowledge Diamond Light Source for time at beamline I05 under proposal SI20259. \\

\textbf{Competing interests} \\
The authors declare no competing interests.\\

\textbf{Authors contributions}\\
V.G., R.Fi., and A.V. grew and prepared single crystals. L.D. and Y.X. performed thermopower measurements. S.J. and M.H. detwinned single crystals. M.H., Q.W., S.J., and R.Fr. carried out x-ray and Laue diffraction measurements.  M.H., Q.W., K.P.K., Y.S., D.S., A.B., and J.C. prepared and carried out the ARPES experiment with the assistance of T.K.K., C.C., J.E.R., P.L.F., F.B., N.C.P., and M.S. M.H. analysed the ARPES data. M.H., M.H.F., and J.C. developed the tight-binding model. M.H., D.S., and J.C. conceived the project. 
All authors contributed to the manuscript. \\

\textbf{Additional information}\\
Correspondence to: M.~Horio (mhorio@issp.u-tokyo.ac.jp) and J.~Chang (johan.chang@phy\-sik.uzh.ch).


\end{document}